\documentclass[conference]{IEEEtran}
\IEEEoverridecommandlockouts

\usepackage{cite}
\usepackage{amsmath,amssymb,amsfonts}
\usepackage{algorithmic}
\usepackage{graphicx}
\usepackage{textcomp}
\usepackage{xcolor,colortbl}
\usepackage{multirow}
\usepackage{hyperref}
\usepackage{float}
\definecolor{greenC}{rgb}{0.7,1,0.7}
\newcommand{\greenC}{\cellcolor{greenC}}  

\begin{document}

\title{Ultrasound Aberration Correction based on \\Local Speed-of-Sound Map Estimation
%\thanks{Funding provided by Swiss National Science Foundation and Innosuisse.}
}
 
\author{\IEEEauthorblockN{Richard Rau,
Dieter Schweizer,
Valery Vishnevskiy, and
Orcun Goksel}
\IEEEauthorblockA{Computer-assisted Applications in Medicine,
ETH Zurich, Switzerland}
}

\maketitle
\begin{abstract}
For beamforming ultrasound (US) signals, typically a spatially constant speed-of-sound (SoS) is assumed to calculate delays. 
As SoS in tissue may vary relatively largely, this approximation may cause wavefront aberrations, thus degrading effective imaging resolution. In the literature, corrections have been proposed based on unidirectional SoS estimation or computationally-expensive a posteriori phase rectification. 
In this paper we demonstrate a direct delay correction approach for US beamforming, by leveraging 2D spatial SoS distribution estimates from plane-wave imaging. 
We show both in simulations and with \textit{ex vivo} measurements that resolutions close to the wavelength limit can be achieved using our proposed local SoS-adaptive beamforming, yielding a lateral resolution improvement of 22\% to 29\% on tissue samples with up to 3\% SoS-contrast (45\,m/s).
We verify that our method accurately images absolute positions of tissue structures down to sub-pixel resolution of a tenth of a wavelength, whereas a global SoS assumption leads to artifactual localizations.

\end{abstract}


\section{Introduction}

In medical ultrasound imaging, typically a constant speed-of-sound is assumed to compute the delays for beamforming. 
In heterogeneous tissue structures this condition of a global SoS is, however, generally not fulfilled. 
For instance, when imaging the liver through fat and muscle tissue, SoS values may vary up to 10\%~\cite{goss_comprehensive_1978}.
Such heterogeneous SoS distributions cause wavefront aberrations, which degrade the resolution and accuracy of any ultrasound imaging modality, due to imprecise delay computations in beamforming; thus also limiting application of US imaging for overweight patients.

In photoacoustics such imprecise delays are typically corrected before beamforming based on the coherence of received signals~\cite{yoon_enhancement_2012}.
A similar approach has been recently proposed for US imaging, however only with a unidirectional (axial) SoS estimation~\cite{ali_distributed_2018}, which has limited use for cases where a large lateral SoS gradients are encountered.
A different approach was proposed in~\cite{jaeger_full_2015}, where SoS estimates are used to adjust the images a\,posteriori by first directionally filtering the beamformed RF data at multiple angles and subsequently correcting the phase locally within each filtered frame. This nevertheless is computationally expensive.

In the recent years, several methods have been proposed that allow for the estimation and 2D mapping of local SoS distribution with conventional US systems, e.g. by time-of-flight recordings together with a passive acoustic reflector~\cite{sanabria_speed--sound_2018} or from minute misalignments between images acquired at different plane-wave angles~\cite{sanabria_spatial_2018,jaeger_computed_2015}. 

In this paper we propose to utilize such 2D spatial SoS distribution estimates for correcting the delays used in US beamforming (Fig.\,\ref{fig:sospipe}a), such that a high resolution image as well as accurate localization of tissue structures is achieved, even in scenarios with heterogeneous SoS distributions. 

\section{Methods}
\subsection{Estimation of the local Speed-of-Sound distribution}
\begin{figure}[t]
\centering
    \includegraphics[width=.5\textwidth]{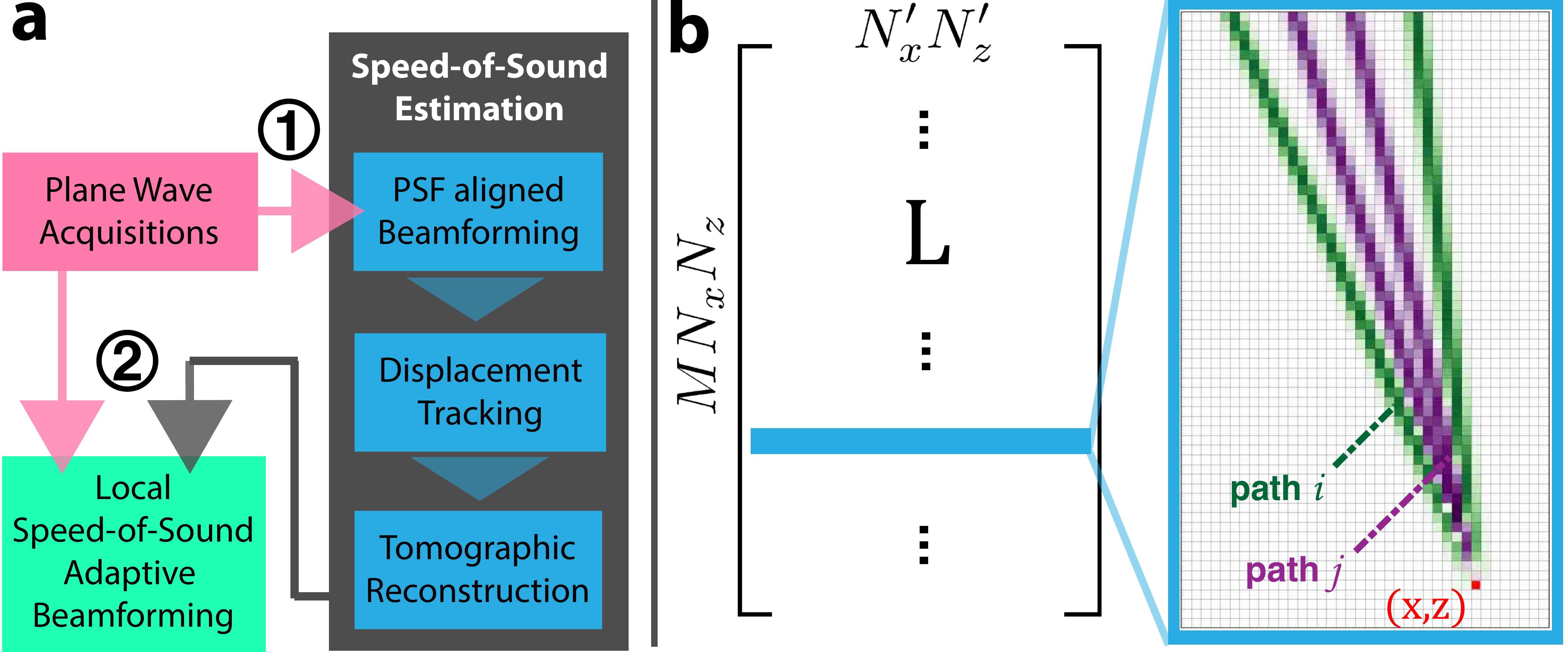}
\caption{(a)~Concept of local speed-of-sound-adaptive beamforming.
    (b)~Illustration of a sample row of the differential path matrix \textbf{L}.
    Each path $i,j$ is composed of a transmit path (determined by plane wave angle) and a receive path (determined by the receive aperture) to/from the pixel $(x,z)$.
    } \label{fig:sospipe}
\end{figure}

2D SoS mapping herein is based on~\cite{sanabria_spatial_2018}, with the following substantial changes that enable improved reconstructions that were not attainable earlier.

We first beamform multiple angled plane waves, assuming a homogeneous (global) SoS $c_0 = 1/\sigma_0$ with slowness $\sigma$, and subsequently estimate apparent motion between some of these beamformed plane wave images in the RF domain using normalized cross-correlation as in~\cite{sanabria_spatial_2018}.
Wavefronts arriving at a tissue location from different plane-wave angles may thus propagate through tissue regions of differing SoS values. 
Information about the SoS distribution is hence encoded at each tissue location in the axial apparent displacements between multiple angled beamformed images.
This allows for a tomographic reconstruction of the slowness $\sigma\in\mathbb{R}^{N_x'N_z'}$ on a $N_x'\times N_z'$ spatial grid, by formulating the inverse problem
\begin{equation}\label{eq:sosrecon}
 \boldsymbol{\hat\sigma} = \arg \min_{\boldsymbol{\sigma}}
 \| \textbf{L}(\boldsymbol{\sigma-\sigma_0}) - \boldsymbol{\Delta\tau} \|_1  +  \lambda \|\textbf{D}\boldsymbol{\sigma} \|_1.\ \ 
\end{equation}
Measured delays $\boldsymbol{\Delta\tau}$$\in$$\mathbb{R}^{M N_x N_z}$ are obtained from the  $M$ different  combinations of displacement tracked images (which are on the $N_x\times N_z$ beamforming grid). 
The differential path matrix $\textbf{L}$$\in$$\mathbb{R}^{M N_x N_z\times N_x' N_z'}$ then links the slowness distribution to the measured delays, e.g. in Fig.\,\ref{fig:sospipe}b to the delay measurement at pixel $(x,z)$ of the displacement tracked combination between the beamformed plane wave images $j$ and $i$.
To overcome the ill-conditioning of $\mathbf{L}$ in~(\ref{eq:sosrecon}), spatial smoothness regularization is applied by matrix $\mathbf{D}$ weighted by $\lambda$, implementing anisotropic weighting of horizontal, vertical and diagonal gradients for limited-angle computed tomography specific regularization to suppress streaking artifacts orthogonal to missing (lateral) projections. 
We assume straight ray propagation and use  $\ell_1-\ell_1$ cost term as seen in~(\ref{eq:sosrecon}) for robust solutions, similar to previous work in this field~\cite{sanabria_spatial_2018,sanabria_speed--sound_2018,Rau_attenuation_19}.
We empirically set $\lambda = 6.5\cdot 10^{-2}$ for all experiments and use an unconstrained optimization package \texttt{minFunc}\footnote{\href{https://www.cs.ubc.ca/~schmidtm/Software/minFunc.html}{https://www.cs.ubc.ca/$\sim$schmidtm/Software/minFunc.html}} to numerically solve~(\ref{eq:sosrecon}).

We herein extend the SoS estimation method in~\cite{sanabria_spatial_2018} by aligning the point-spread-function (PSF) of the beamformed RF signals~\cite{stahli_forward_2019}. When beamforming differently angled plane waves with the same receive apertures, the respective PSF angles are dependent on the plane wave angle. 
Especially for strongly scattering point-like objects, this may introduce false motion estimation due to the PSF misalignment.
To suppress such artifacts, similarly to~\cite{stahli_forward_2019}, we adapt the receive apertures such that the PSFs of two beamformed angled plane wave images to be displacement tracked are aligned at a defined angle $\theta_{PSF}$.
The axial displacement values are furthermore corrected by projection onto the minor (modulation) axis of the above-aligned PSF and scaled by an empirically calibrated factor of 1.5 for higher reconstruction accuracy.

\subsection{Local Speed-of-Sound-Adaptive Beamforming}

Delays $\boldsymbol{\tau}$$\in$$\mathbb{R}^{N_cN_xN_z}$ from all pixels on the $N_x\times N_z$ beamforming grid to all $N_c$ transducer elements can be computed given $\boldsymbol{\hat\sigma}$ from SoS image reconstruction using
\begin{equation}\label{eq:dassos}
 \boldsymbol{\tau} = \mathbf{P}\boldsymbol{\hat\sigma}
\end{equation}
with the path matrix $\mathbf{P}$$\in$$\mathbb{R}^{N_cN_xN_z\times N_x'N_z'}$ that discretizes time-of-arrivals on a Cartesian grid and is similar to $\mathbf{L}$ in~(\ref{eq:sosrecon}) with the difference that $\mathbf{P}$ contains a single path per row instead of four paths in $\mathbf{L}$ seen in Fig.\,\ref{fig:sospipe}b.

We then beamform the separate acquired plane-wave raw RF data with the corrected delays $\boldsymbol{\tau}$ from~(\ref{eq:dassos}) using conventional delay-and-sum
\begin{equation}\label{eq:das_conv}
s(x,z) = \sum_{n_c=1}^{N_c}A(n_c,x,z)\cdot RF(n_c,\tau(n_c,x,z)),
\end{equation}
where $A(n_c,x,z)$$\in$$[0,1]$, denotes dynamic aperture with apodization. 
Hereafter, beamforming using corrected delays is referred as \textit{local speed-of-sound-adaptive beamforming}. 
We use (\ref{eq:dassos} - \ref{eq:das_conv}) also for the initial beamforming with homogeneous global SoS, where $\boldsymbol{\hat\sigma}=\sigma_0 = 1/c_0$ is a constant.

\subsection{Experiment Design}
We evaluated our delay correction approach using two experiments below, one simulation and one \textit{ex vivo} study.

\begin{figure}[htbp]
\centering
    \includegraphics[width=.48\textwidth]{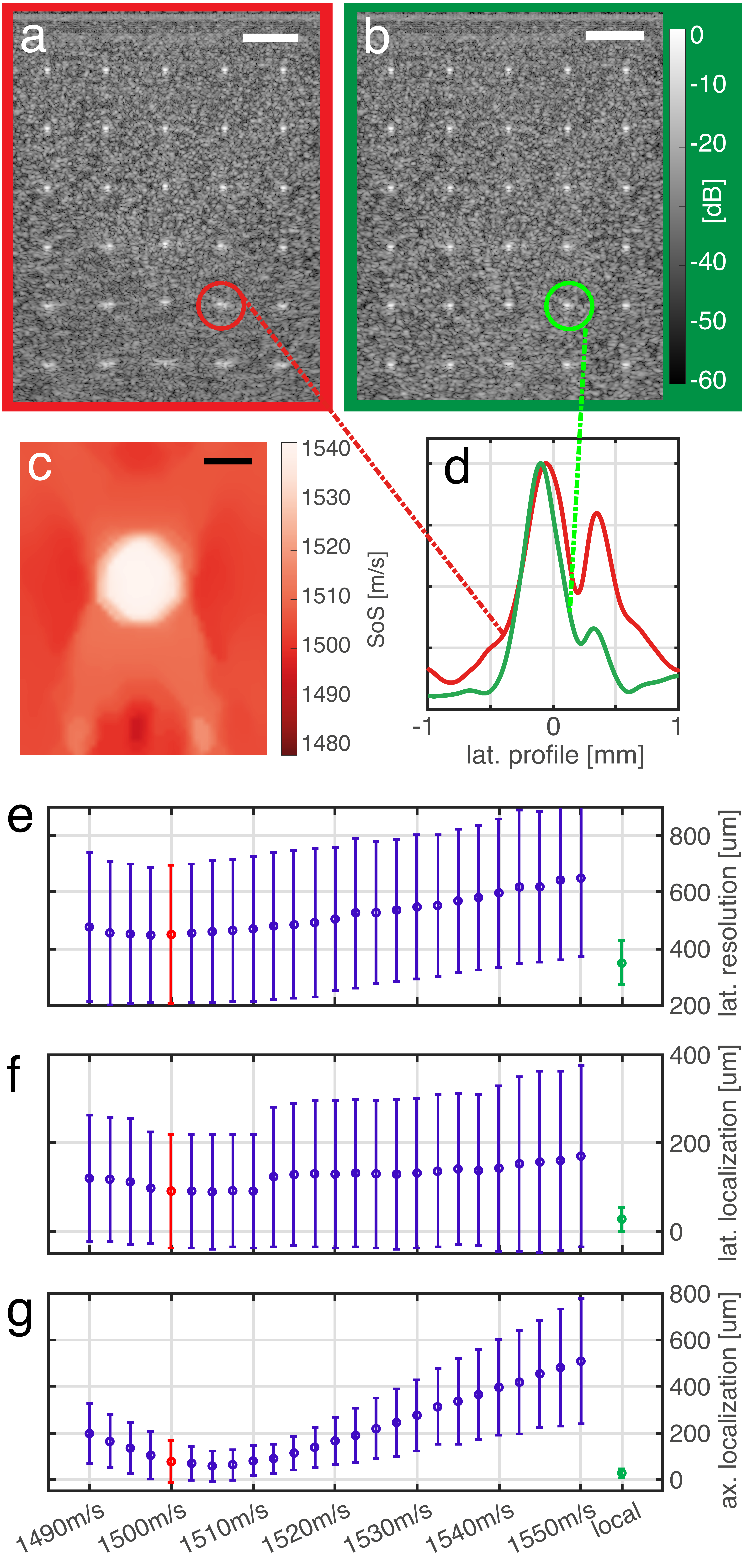}
    \caption{\textbf{Simulation: Local SoS-adaptive vs. global SoS beamforming.} 
     (a)~The B-Mode image beamformed at a global speed-of-sound of $1500\,\mathrm{m/s}$ yielding the best resolution within a range, and (b)~the B-Mode image obtained using our local SoS-adaptive beamforming. 
     (c)~The reconstructed SoS distribution that was used in local SoS-adaptive beamforming.
     (d)~A sample lateral profile of a point scatterer in linear scale shows strong aberration effects for the global SoS beamforming (red), which is alleviated with the local SoS-adaptive case~(green). 
     (e)~Resolution degradation, (f)~lateral localization error and (g)~axial localization error caused by the aberration effects.
     The blue and red errorbars in (e-g) indicate the mean and standard deviations of the metrics for all scatterers for beamforming with various global SoS values; red is the best-resolution case at $1500\,\mathrm{m/s}$, with the B-mode image shown in~(a).
     The green errorbar indicates our local SoS-adaptive beamforming. 
     All scale-bars represent a length of $5mm$.
    } \label{fig:kwaveresults}
\end{figure}

\noindent {\bf Simulation: } 
We used k-wave~\cite{treeby_k-wave:_2010} for a medium discretized on a 75um resolution grid.
SoS heterogeneity is modeled as a $10\,\mathrm{mm}$ diameter circular inclusion of $1545\,\mathrm{m/s}$ on a background substrate of $1500\,\mathrm{m/s}$ (i.e.\ 3\% contrast) (see Fig.\,\ref{fig:kwaveresults}c).
For evaluation of resolution/localization error, a $ 5\times 6 $ point scatterer grid is created by increasing the medium density at the corresponding pixels (cf. Fig.\,\ref{fig:kwaveresults}b).
A realistic speckle pattern is realized by increasing a random 10\% set of the medium pixels by a slight perturbation in density value.

\noindent {\bf  \textit{ex vivo} study: } 
We used a porcine skeletal muscle sample and a strong SoS contrast was achieved by removing a cylindrically shaped part of the tissue from the center (radius $\approx 3mm$) and replacing it with a gelatin/water mix (10\% gelatin in water per weight), see Fig.\,\ref{fig:exvivoresults}a.
For the resolution analysis a $7\times 3$ point scatterer grid was created below the implanted SoS inclusion, using metal wires of $250\,\mathrm{\mu m}$ diameter.
Unbeamformed RF data was acquired using a Fukuda Denshi UF-760AG ultrasound system.

\subsection{Parameters for Acquisition and Processing}

In both experiments above, RF data was acquired at $5\,\mathrm{MHz}$ center frequency and for plane waves with an angle range of $\pm25^\circ$. 
For the SoS estimation only a subset of angles was used, i.e.\ [$-12^\circ,-10^\circ,-8^\circ,$  $... $ $,12^\circ$], because large angles were found to lead to suboptimal displacement tracking results.
Each of these plane waves were beamformed using PSF centering for three different PSF angles; $[-15^\circ,0^\circ,15^\circ]$.

For the beamforming evaluation comparing our local SoS-adaptive beamforming to a global SoS assumption, 11 plane waves angled at [$-25^\circ,-20^\circ,-15^\circ,$  $... $ $,20^\circ$] were separately beamformed using (\ref{eq:das_conv}) and coherently compounded.

\section{Results}
\subsection{Simulation Study}

The results of the simulation study are summarized in Fig.\,\ref{fig:kwaveresults}.
The B-Mode images show qualitatively two improvements with the local SoS-adaptive beamforming (Fig.\,\ref{fig:kwaveresults}b) compared to the global SoS case (Fig.\,\ref{fig:kwaveresults}a):
1)~A distinct speckle pattern is maintained within the whole field-of-view and 2)~the point scatterers at the deeper locations below the SoS-inclusion (cf. Fig.\,\ref{fig:kwaveresults}c) have a sharper appearance and are better resolved. 
With the global SoS assumption, the image quality degrades due to wavefront aberrations that cause incoherent summation of the delayed receive signals.
This results in inferior lateral PSF (envelope) profile as seen in Fig.\,\ref{fig:kwaveresults}d for a representative point scatterer. 

For a quantitative assessment, we evaluated the lateral resolution as well as the correct localization of point scatterers placed across the imaging region (Fig.\,\ref{fig:kwaveresults}e-g). 
The resolution is computed as the full-width-at-half-maximum of the PSF envelope.
Localization errors were quantified as the axial and lateral distance from the location of PSF envelope maximum to the known ground-truth scatterer positions in the simulation.
To test the hypothesis whether any single global sound-speed value could perform superior to our beamforming with local SoS mapping, we evaluated all the three metrics given above for multiple global SoS values ranging from $1490\,\mathrm{m/s}$ to $1550\,\mathrm{m/s}$. 
As seen in Fig.\,\ref{fig:kwaveresults}e-g, optimality definition of a global SoS value may depend on the chosen metric (e.g., optimal lateral resolution is at a lower SoS than the optimal axial localization). 
Nevertheless, regardless of the metric choice, our proposed local SoS-adaptive beamforming is substantially superior to any conventional global-SoS beamforming, cf.\ Tab.\ref{tabresults}. 
With our method, average lateral resolution improved by over 22\% to $(349\pm79)\,\mathrm{\mu m}$, which is close to the wavelength ($\approx 300\,\mathrm{\mu m}$) of the US pulse. 
Scatterer localization accuracy improved significantly by up to 70\% with an average localization error of $28\,\mathrm{\mu m}$ in both axes, which indicates sub-pixel resolution given the beamforming grid~(i.e.\ $37.5\mu m \times 75\mu m$).

\subsection{\textit{ex vivo} Study}
\begin{figure}[t]
\centering
    \includegraphics[width=.48\textwidth]{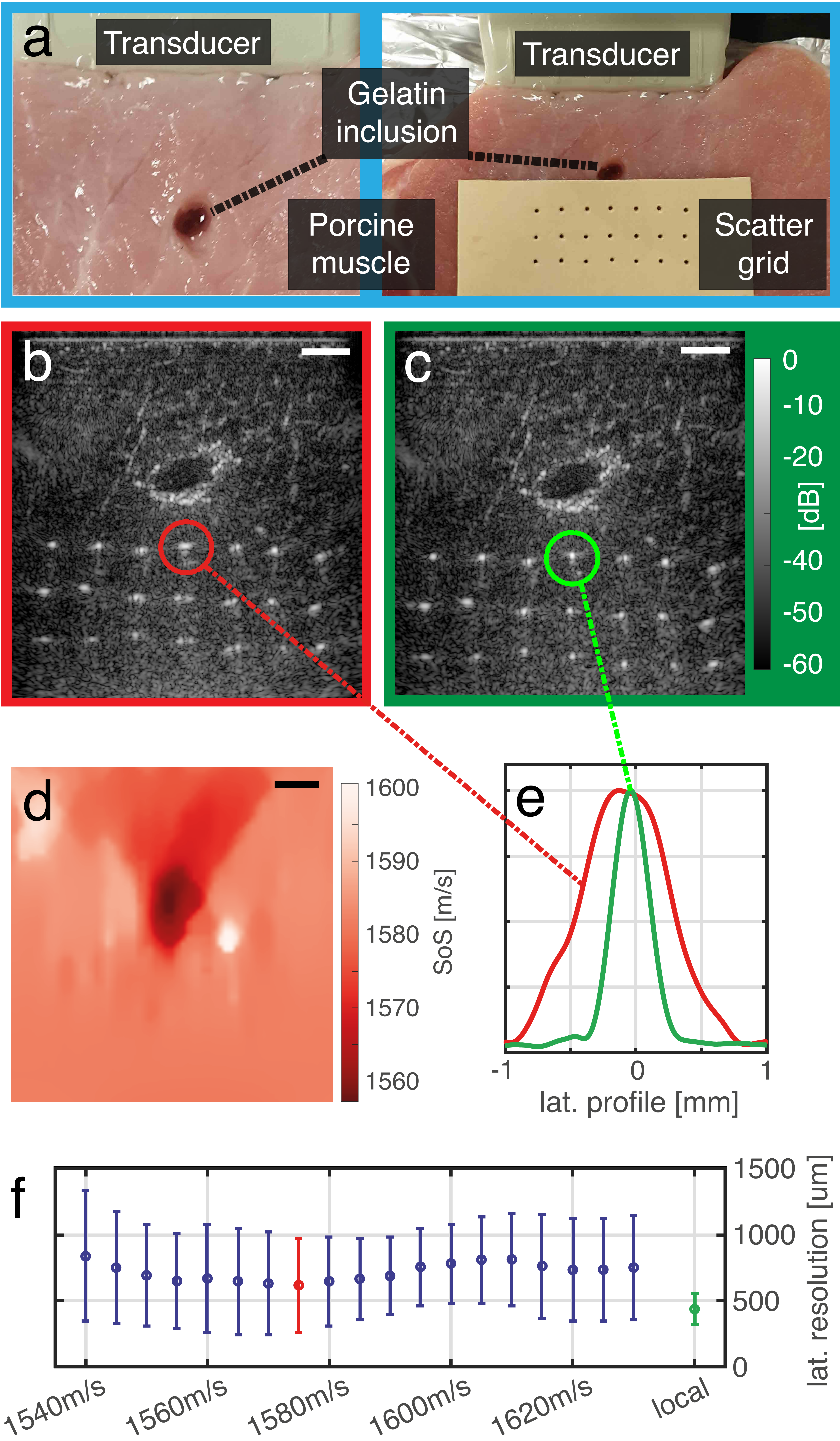}
    \caption{\textbf{\textit{ex vivo}: Local SoS-adaptive vs.\ global SoS beamforming.} 
    (a)~The porcine skeletal muscle sample with a gelatin inclusion. 
    The scatter grid on the right indicates where the $250\,\mathrm{\mu m}$ metal wires were placed. 
    (b\&c)~B-mode images showing the hypoechoic gelatin inclusion and the point scatterers. 
    (d)~2D SoS reconstruction where the gelatin demonstrates a considerably lower SoS compared to the surrounding muscle tissue. 
    The B-Mode image in (b) is beamformed with a global SoS $1575\,\mathrm{m/s}$) and in (c) with the local SoS-adaptive method.  
    (e)~The latter also leads to a narrower Gaussian shaped PSF envelope in the lateral axis (in linear scale). 
    Similarly to the simulation results in Fig.\,\ref{fig:kwaveresults}, the best lateral resolution is achieved with the local SoS-adaptive method~(f). All scale-bars represent a length of $5mm$.  
      } \label{fig:exvivoresults}
\end{figure}

The \textit{ex vivo} study results are summarized in Fig.\,\ref{fig:exvivoresults}.
Similarly to the simulation results, strong aberration effects are visible in the B-Mode image when a global SoS is assumed (Fig.\,\ref{fig:exvivoresults}b). 
The aberration effects are caused by the heterogeneous SoS distribution (Fig.\,\ref{fig:exvivoresults}c), which is herein introduced by the gelatin inclusion ($1560\,\mathrm{m/s}$ gelatin vs.\ $\approx$1585\,$\mathrm{m/s}$ porcine skeletal muscle).
These aberration effects are corrected by our local SoS-adaptive method as can be seen from the B-Mode image in Fig.\,\ref{fig:exvivoresults}c as well as from the sample lateral profile in Fig.\,\ref{fig:exvivoresults}e.
For the global SoS assumption, the best achievable resolution is at $1575\,\mathrm{m/s}$ as seen in Fig.\,\ref{fig:exvivoresults}f, which is improved by over 29\% to $(437\pm 118)\,\mathrm{\mu m}$ by our local SoS-adaptive beamforming as tabulated in Tab.\ref{tabresults}.

\begin{table}[htbp]
\caption{Results of the resolution and localization error analysis}
\begin{center}
\begin{tabular}{|l|l|c|c|}
\cline{3-4} 
\multicolumn{2}{c|}{} & \textbf{optimal}             & \textbf{local SoS} \\
\multicolumn{2}{c|}{} & \textbf{global SoS}            &  \textbf{adaptive} \\
\hline
\multirow{3}{*}{\textbf{Simulation}} & lat. resolution $[\,\mathrm{\mu m}]$      & $448\pm245$     & \greenC $349\pm79$  \\
\cline{2-4} 
                     & lat. localization error $[\,\mathrm{\mu m}]$     & $90\pm132$      &  \greenC $28\pm27$ \\
\cline{2-4} 
                     & ax. localization error $[\,\mathrm{\mu m}]$       & $58\pm67$       & \greenC $28\pm20$ \\
\hline
\hline
\textit{\textbf{ex vivo}}     & lat. resolution $[\,\mathrm{\mu m}]$      & $617\pm360$      & \greenC $437\pm118$\\
\hline
\end{tabular}
\label{tabresults}
\end{center}
\end{table}

\section{Discussion and Conclusions}

In this paper we have presented a novel method on how 2D speed-of-sound maps can be used for improving beamforming of medical ultrasound images. 
This is achieved by correcting (calculating) beamforming delays based on the SoS reconstruction. 
It is shown that with our proposed method a lateral resolution close to the wavelength limit can be achieved, with improvements of more than 22\% in simulation and of more than 29\% in \textit{ex vivo} experiments. 
The axial resolution was not evaluated, because no significant degradation was observed in either case. 

The beamforming accuracy was furthermore analyzed in terms of correctness with respect to the physical locations, using a grid of imaged point scatterers. 
With the local SoS-adaptive method, localization errors were minimal with $28\,\mathrm{\mu m}$ on average, which is smaller than the spatial grid resolution.
The assumption of a global homogeneous SoS in beamforming led in this case to a 2-to-3 times higher localization error.

In this study, the largest SoS contrast was 3\% ($45\,\mathrm{m/a}$), which is on the order of maximum variations expected in the breast~\cite{Ruby_breast_19}.
Nevertheless, higher SoS contrasts, e.g.\ of up to or ore than 10\%, may be expected between other tissues such as muscle and fat tissues. 
Thus, further relative improvements from using local SoS-adaptive beamforming can be expected in such scenarios.

In this paper, the improvements of the beamforming were analyzed for B-Mode imaging, nevertheless, it is similarly applicable to other US imaging modalities where tissue with heterogeneous SoS distribution is encountered. 
Smaller PSFs could naturally affect other derived imaging modalities as well, such as displacement tracking in elastography.
A further example would be the imaging of small vessels through a bone layer, which is for instance a problem in functional US imaging~\cite{rau_3d_2018}.

A practical limitation for a real-time implementation of the presented method is the time-consuming algebraic reconstruction employed for mapping local SoS. 
A variational network solution similar to~\cite{vishnevskiy_deep_2019} with inference times on the order of milliseconds could help to overcome this limitation towards real-time speed-of-sound corrected imaging.

\vspace{3ex}{\bf Funding} was provided by the Swiss National Science Foundation and Innosuisse.

\bibliographystyle{IEEEtran}
\bibliography{refs}

\end{document}